\documentclass{aa}  

\usepackage{graphicx}   
\usepackage{amsmath}    
\usepackage{amsmath, amsfonts, amssymb, graphics,graphicx, wrapfig, nicefrac}
\usepackage{epsfig}
\usepackage{epstopdf}
\usepackage{multirow}
\usepackage{multicol}
\usepackage{graphicx}
\usepackage{rotating}
\usepackage{mathrsfs}
\graphicspath{{./}{figures/}}
\usepackage{txfonts}
\begin{document}

   \title{A new measurement of the Galactic $^{12}$C/$^{13}$C gradient from sensitive HCO$^+$ absorption observations}


   \author{Gan Luo
          \inst{1}
          \and
          Laura Colzi\inst{2}
          \and
          Tie Liu\inst{3}
          \and
          Thomas G. Bisbas\inst{4}
          \and
          Di Li\inst{5,6}
          \and
          Yichen Sun\inst{7}
          \and
          Ningyu Tang\inst{8}
          }

   \institute{Institut de Radioastronomie Millimetrique, 300 rue de la Piscine, 38400, Saint-Martin d’Hères, France\\
            \email{luo@iram.fr}
        \and
             Centro de Astrobiología (CAB), CSIC-INTA, Ctra. de Ajalvir Km. 4, 28850 Torrejón de Ardoz, Madrid, Spain
        \and
            Shanghai Astronomical Observatory, Chinese Academy of Sciences, 80 Nandan Road, Shanghai 200030, China
        \and
            Research Center for Astronomical Computing, Zhejiang Laboratory, Hangzhou 311100, China
        \and
             Department of Astronomy, Tsinghua University, Beijing 100084, China
        \and
             CAS Key Laboratory of FAST, National Astronomical Observatories, Chinese Academy of Sciences, Beijing 100101, China
        \and
            School of Astronomy and Space Science, Nanjing University, Nanjing 210093, China
        \and
            Department of Physics, Anhui Normal University, Wuhu, Anhui 241002, China
             }

   \date{Received xx; accepted xx}

\abstract{We present a new constraint on the Galactic $^{12}$C/$^{13}$C gradient with sensitive HCO$^+$ absorption observations against strong continuum sources. The new measurements suffer less from beam dilution, optical depths, and chemical fractionation, allowing us to derive the isotopic ratios precisely. The measured $^{12}$C/$^{13}$C ratio in the solar neighborhood (66$\pm$5) is consistent with those obtained from CH$^+$. Two measurements toward the GC are 42.2$\pm$1.7 and 37.5$\pm$6.5. Though the values are a factor of two to three higher than those derived from dense gas tracers (e.g., H$_2$CO, complex organic molecules) toward Sagittarius (Sgr) B2 regions, our results are consistent with the absorption measurements from c-C$_3$H$_2$ toward Sgr B2 ($\sim$40) and those from CH$^+$ toward Sgr A$^*$ and Sgr B2(N) ($>$30). We have calculated a new Galactic $^{12}$C/$^{13}$C gradient of (6.4$\pm$1.9)$R_{\rm GC}$/kpc+(25.9$\pm$10.5) and found an increasing trend of the $^{12}$C/$^{13}$C gradient obtained from high-density to low-density gas tracers, suggesting that opacity effects and chemical fractionation may have a strong impact on the isotopic ratios observed in high-density regions. }
 


   \maketitle
%

\section{Introduction}

The measurements of isotopic ratios in the interstellar medium (ISM) are important for understanding Galactic chemical evolution (GCE) due to different stellar yields with different masses \citep{Guesten1982,Maeder1983,Iben1983,Clayton2003,Kobayashi2011,Romano2017,Kobayashi2020}. One of the few isotopic ratios that can be easily measured is $^{12}$C/$^{13}$C . Notably, $^{12}$C is mainly produced through He-burning as the primary element (e.g., core-collapse supernovae and He-burning shells in low-mass stars), while $^{13}$C is an intermediate product of CNO cycling in intermediate-mass asymptotic giant branch (AGB) stars \citep{Kobayashi2011,Romano2017}. Since the timescale in which massive stars release the elements into the ISM is much shorter than for low-to-intermediate mass stars, the $^{12}$C/$^{13}$C ratios predicted from GCE models would decrease as a function of time with a given initial mass function (IMF).  
In the Milky Way, a radial gradient in the $^{12}$C/$^{13}$C ratio inferred from $^{12}$C$^{18}$O and $^{13}$C$^{18}$O rises from approximately 24 at a galactocentric distance ($R_{\rm GC}$) of $\sim$ 0.5 kiloparsec (kpc) to roughly 70 at $R_{\rm GC}$ $\sim$ 12 kpc \citep{Langer1990,Langer1993}. 

While various methodologies and tracers have been employed in recent years to constrain the Galactic $^{12}$C/$^{13}$C gradient \citep{Halfen2017,Yan2019,Jacob2020,Sun2024}, the different methods have yielded inconsistent results, especially for the Galactic center (GC) regions. 
Measuring the $^{12}$C/$^{13}$C isotopic ratio with molecular lines faces challenges from both observational factors (e.g., opacity, beam dilution) and chemical effects (e.g., selective photodissociation and fractionation). The most abundant carbon-bearing isotopologs (e.g., $^{12}$CO, H$^{12}$CO$^+$, H$^{12}$CN) can easily become optically thick in dense clouds, making precise estimation of column densities difficult. In diffuse and translucent clouds, the main isotopologs may not pose challenges for optical depths, but the rare isotopologs are often difficult to detect due to their low abundance and sub-thermally excitation of even the ground state transition \citep{Godard2010,Luo2020}. Additionally, distant clouds have smaller beam-filling factors, which reduces the reliability of the measurements observed with a single dish. 

Isotope-selective photodissociation and isotopic exchange reactions can strongly alter the $^{12}$C/$^{13}$C isotopic ratios in molecules from the original element ratios. For abundant species such as CO, the self-shielding of $^{12}$CO is significantly stronger than less abundant isotopologs (e.g., $^{13}$CO and C$^{18}$O). Thus, the $^{12}$C/$^{13}$C ratio in CO would be higher under the existence of external UV radiation \citep{Bally1982,Chu1983,Liszt2007}. Additionally, due to the differences in zero-point energy between different isotopologs, isotopic exchange reactions can further alter $^{12}$C/$^{13}$C ratios in various molecules. For instance, the isotopic exchange reaction between $^{13}$C$^+$ and CO is exothermic, leading to an enrichment of $^{13}$C in CO and other molecules primarily formed through CO in cold environments \citep{Watson1976,Langer1984,Roueff2015,Colzi2020,Sipila2023}.

Absorption lines against strong continuum sources can overcome such difficulties. The advantage of using absorption lines is to measure the optical depth as well as the column density precisely even if the transition is sub-thermally excited. Such a methodology has been widely used to quantify the physical properties of low-density gas \citep{Lucas1996,Lucas1998,Luo2020,Luo2023b,Rybarczyk2022a,Liszt2023,Gerin2024}. \citet{Lucas1998} is one of the first works that measured the isotopic ratios in the solar neighborhood through millimeter molecular absorption lines. \citet{Liszt2018} measured the H$^{12}$CO$^+$/H$^{13}$CO$^+$ ratios in four directions toward the Galactic bulge, and they inferred a much higher value ($\sim$60) than the other tracers ($\sim$20). Inspired by these pioneering works, systematic measurements at different $R_{\rm GC}$ with such a methodology could be useful to constrain the $^{12}$C/$^{13}$C gradient.

In this work, we present a new measurement of the Galactic $^{12}$C/$^{13}$C gradient by using absorption lines of H$^{12}$CO$^+$ and H$^{13}$CO$^+$ with high-angular observations of the Atacama Large Millimeter/submillimeter Array (ALMA) and the Northern Extended Millimeter Array (NOEMA). We emphasize that such a methodology to constrain the isotopic gradient of the Milky Way with a large sample, especially for future searches in the anti-center direction (though rare and challenging), could provide valuable constraints for GCE models.

\section{Observations}\label{sec:obs}

\subsection{Quasar sight lines}\label{sec:obs_qso}
\begin{figure}
\centering
\includegraphics[width=0.95\linewidth]{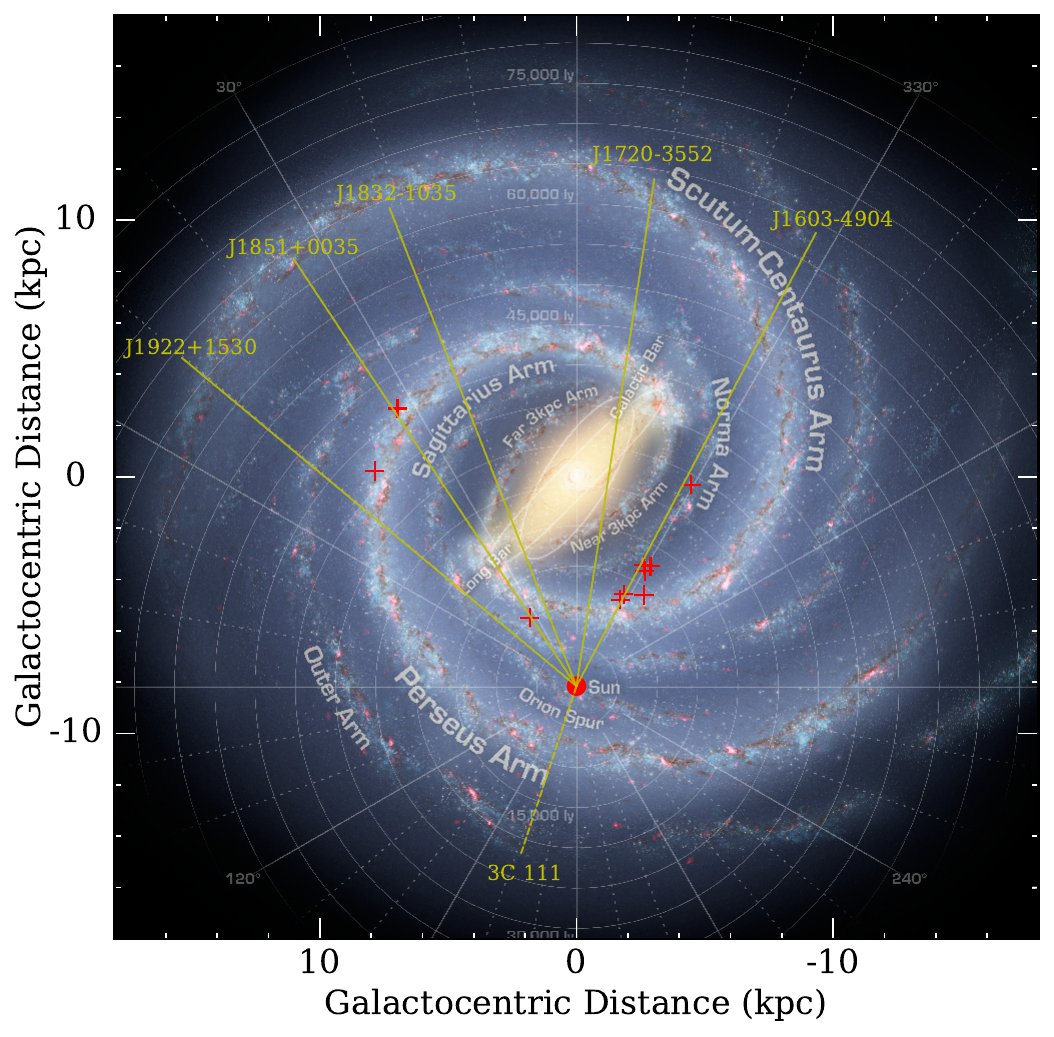}
\caption{Distribution of the observed quasar sight lines (yellow lines) and ultra-compact H\,{\sc ii} region (red crosses) projected onto a top-down schematic view of the Milky Way (artist’s concept, R. Hurt: NASA/JPLCaltech/SSC). \label{fig:srn}}
\end{figure}

We carried out H$^{12}$CO$^+$ and H$^{13}$CO$^+$ {\it J}=1-0 observations toward six quasars (Fig.~\ref{fig:srn}). Five were observed with ALMA (2022.1.01438.S, PI: Thomas G. Bisbas) during March 20--30, 2023, and one source (3C~111) was observed with NOEMA (W20BB, PI: Gan Luo) in December 2020 and January 2021. The angular resolution of ALMA is $\sim$0.6$''$ at 89\,GHz. 
The spectral resolution of ALMA is 122\,kHz, corresponding to a velocity resolution of $\sim$0.4\,km\,s$^{-1}$ at 89\,GHz. The raw data was calibrated using the standard pipeline with the Common Astronomy Software Applications \citep[CASA, version 6.4.1;][]{CASA2022}. The imaging of the calibrated visibilities was performed using the {\it tclean} algorithm with Briggs weighting (robust = 0.5). 
The NOEMA observations have an angular resolution of $\sim$2.1$''$ at 89\,GHz and a velocity resolution of $\sim$0.21\,km\,s$^{-1}$. The PolyFiX backend of NOEMA covers H$^{12}$CN, H$^{13}$CN, HN$^{12}$C, and HN$^{13}$C {\it J}=1-0 transitions simultaneously. The calibration of the raw data was performed using {\sc clic} software in the {\sc gildas}.\footnote{https://www.iram.fr/IRAMFR/GILDAS} 
The typical optical depth noise levels are 3$\sim$7$\times$10$^{-3}$ per 0.4 km\,s$^{-1}$ velocity channel for ALMA observations and 2$\times$10$^{-3}$ per 0.21 km\,s$^{-1}$ velocity channel for NOEMA observations.

\subsection{UCH\,{\sc ii} sight lines}\label{sec:obs_uchii}
The ALMA observations toward ten UCH\,{\sc ii} regions (red crosses in Fig.~\ref{fig:srn}), which were selected from the ALMA Three-millimeter Observations of Massive Star-forming regions (ATOMS) survey (2019.1.00685.S, PI: Tie Liu), were observed from September to November 2019 with both 12-m and 7-m arrays at band 3. The baselines of ATOMS range from 15 to 783.5\,m, corresponding to an angular resolution of $\sim$1.2$''$ at 89\,GHz. The spectral resolution of H$^{12}$CO$^+$ and H$^{13}$CO$^+$ {\it J}=1-0 transitions are 0.2\,km\,s$^{-1}$ and 0.422\,km\,s$^{-1}$, respectively. The raw data were calibrated using CASA version 5.6, and the combination and imaging of 12-m and 7-m data were performed with CASA version 6.4.1. The sensitivity of the resultant spectra depends on the properties of each source, a detailed description of observations can be found in \citet{Liu2020}.

\section{Results}\label{sec:results}
\subsection{Absorption spectra and integrated optical depth}\label{sec:spec}

\begin{figure}
\centering
\includegraphics[width=0.95\linewidth]{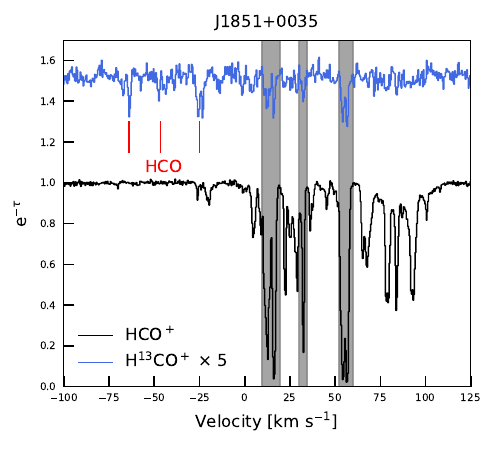}
\caption{Normalized absorption spectra of HCO$^+$ (black curve) and H$^{13}$CO$^+$ (blue curve) toward J1851+0035. The spectrum of H$^{13}$CO$^+$ has been scaled by a factor of five and shifted upward by 0.5 for better display. The gray shaded regions denote the velocity ranges used to calculate the column density ratios. The red vertical lines represent HCO absorption, where the $-25$km\,s$^{-1}$ component of H$^{13}$CO$^+$ absorption is contaminated with HCO {\it J}=1/2-1/2, {\it F}=1-1. Thus, these components were discarded since they can only obtain lower limits of $^{12}$C/$^{13}$C. \label{fig:fig1}}
\end{figure}

The absorption spectra of HCO$^+$ (hereafter, HCO$^+$ refers to H$^{12}$CO$^+$) and H$^{13}$CO$^+$\footnote{All transitions refers to {\it J}=1-0 rotational transition unless otherwise noted.} were extracted from the continuum peak of each source. Figure \ref{fig:fig1} shows an example of normalized spectra of HCO$^+$ and H$^{13}$CO$^+$ toward J1851+0035. Spectra toward all other sources can be found in Appendix \ref{sec:a}. 

Since most of the sources are located in the Galactic plane ($|b| < 2^{\circ}$), the foreground absorption components are mixed, and it is difficult to decompose individual Gaussian components (except for the high-latitude source 3C111; see Appendix \ref{sec:b}). To obtain the column density ratios, we integrated the optical depths ($\int \tau_\nu d\upsilon$) of HCO$^+$ and H$^{13}$CO$^+$ in the same velocity range (shadowed region in Fig.~\ref{fig:fig1} and \ref{fig:allspec}). We set four criteria when identifying the integrated range: 1) H$^{13}$CO$^+$ absorption should not be contaminated with absorption features of other molecules (e.g., $-25$ km\,s$^{-1}$ component of H$^{13}$CO$^+$ in Fig.~\ref{fig:fig1}); 2) each of the integrated ranges of H$^{13}$CO$^+$ should be distinguished from the others (no overlap); 3) the absorption profile of HCO$^+$ should not be saturated ($e^{-\tau}$ > rms); and 4) absorption spectra toward UCH\,{\sc ii} regions should not be contaminated with emission features from the compact foreground envelopes. The uncertainty of $\int \tau_\nu d\upsilon$ ($\sigma_{\rm I}$) through an error propagation formula is thus defined by
\begin{equation}
    \rm \sigma_I = \frac{d\upsilon}{2} \sqrt{\sum_{i=1}^{N-1} (\sigma_{\tau_i}^2+\sigma_{\tau_{i+1}}^2)},
\end{equation}
where ${\rm d\upsilon}$ is the channel width, N is the length of the data array in the velocity range over which the opacity profile is integrated, and ${\rm \sigma_{\tau_i} = RMS/e^{-\tau_i}}$ is the uncertainty of $\tau$ at the $i$th channel.
The integrated optical depths of HCO$^+$ and H$^{13}$CO$^+$ toward all sources are listed in Table \ref{table:tab2}.

The molecular column density can be written as a function of $\int \tau_\nu d\upsilon$ \citep{Mangum2015}:
\begin{equation}
N_{\rm tot} = \frac{3h}{8{\pi}^3\left | \mu_{\rm lu} \right |^2} \frac{Q_{\rm rot}}{g_{\rm u}} \frac{e^{\frac{E_{\rm u}}{kT_{\rm ex}}}} {e^{\frac{h\nu}{kT_{\rm ex}}}-1}  \int \tau_\nu d\upsilon,
\label{eq:n_tot}
\end{equation}
where $h$ is the Planck constant, $k$ is the Boltzman constant, $\left | \mu_{\rm lu} \right |^2$ is the dipole matrix element, $Q_{\rm rot}$ is the rotational partition function, $g_{\rm u}$ is the degeneracy of the upper energy level, $E_{\rm u}$ is the upper energy level, and $T_{\rm ex}$ is the excitation temperature.
For each transition, $\left | \mu_{\rm lu} \right |^2$, $E_{\rm u}$, and the rest frequency $\nu$ were taken from the Cologne Database for Molecular Spectroscopy \citep[CDMS;][]{Muller2001,Muller2005}, and they are listed in Table \ref{tab:transitions}. 
For linear molecules, the simplified partition function is given by \citet{McDowell1987}
\begin{equation}
Q_{\rm tot} = \frac{kT}{hB_0} e^{\frac{hB_0}{3kT}} ,
\label{eq:q_tot}
\end{equation}
where $B_0$ is the rigid rotor rotation constant.

The excitation temperature of HCO$^+$ in diffuse molecular clouds is usually close to the cosmic microwave background temperature ($T_{\rm CMB}$ = 2.73 K) \citep{Godard2010,Luo2020}. If we fill in all the constants in Eq.~\ref{eq:n_tot} and take $T_{\rm ex} = 2.73$\,K for both HCO$^+$ and H$^{13}$CO$^+$, the column density ratios can be simplified as
\begin{equation}
    \frac{N_{\rm HCO^+}}{N_{\rm H^{13}CO^+}} = 0.974 \frac{\int \tau_{\rm HCO^+} d\upsilon}{\int \tau_{\rm H^{13}CO^+} d\upsilon}.
\end{equation}
Though $T_{\rm ex}$ of HCO$^+$ can rise above 2.73 K when the gas density is higher than 300 cm$^{-3}$ \citep{Rybarczyk2022b}, a variance of $T_{\rm ex}$ from 2.73 to 5 K only results in a variance of the column density ratio of less than 1\%. 
The calculated ratios in our samples range from 12.7$\pm$0.2\footnote{However, this low value ($< 30$) is only derived at $V_{\rm lsr} = -22$ km\,s$^{-1}$ toward J1720-3552; see Appendix \ref{sec:e} for more detail.} to 139.4$\pm$38.0 (Table \ref{table:tab1}).

\subsection{The $^{12}$C/$^{13}$C gradient}\label{sec:12_13c_gradient}

Obtaining an accurate distance of a distant molecular cloud is difficult, especially without parallax measurements from maser emissions. The most common estimation of such a distance is based on kinematic distance. We calculated the $R_{\rm GC}$ for each velocity component with the Monte Carlo kinematic distance method, using the latest rotational curve and updated solar motion parameters \citep{Wenger2018}. This model is obtained through the trigonometric parallax results from the high-mass star-forming regions (HMSFRs) in the BeSSeL Survey \citep{Reid2014,Reid2019}. 
This method samples the local standard of rest velocity ($V_{\rm lsr}$) and Galactic rotational curve parameters for HMSFRs and derives the probability density distribution (PDF) of kinematic distance, resulting in a median uncertainty of 13\% to the parallax distances. A detailed description of this method can be found in \citet{Wenger2018}.

The optical depth-weighted $V_{\rm lsr}$ in our samples is calculated by
\begin{equation}
    V_{\rm lsr} = \frac{\int \upsilon \tau_{\rm HCO^+} d\upsilon}{\int \tau_{\rm HCO^+} d\upsilon}.
\end{equation}
The calculated $V_{\rm lsr}$ and $R_{\rm GC}$ as well as the uncertainties are listed in Table \ref{table:tab1}.  We note that such an estimation with pure circular motions may fail to predict the $R_{\rm GC}$ near the GC (e.g., the highly negative $V_{\rm lsr}$ toward J1720-3552), in which non-circular motions should be taken into account \citep{Liszt2018}. Therefore, we used the tilted-disk model presented by \citet{Burton1978} to estimate the velocity components of $V_{\rm lsr} < -140$ km\,s$^{-1}$ toward J1720-3552, resulting in an $R_{\rm GC}$ between 1.24 to 1.5 kpc. Thus, we adopted a value of 1.37$\pm$0.04 kpc for this component.

\begin{figure}
\centering
\includegraphics[width=0.95\linewidth]{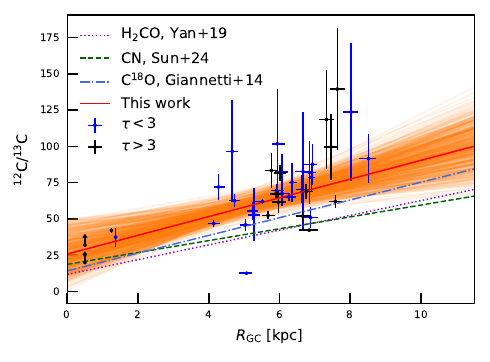}
\caption{Measurements of H$^{12}$CO$^+$/H$^{13}$CO$^+$ as a function of $R_{\rm GC}$ and comparison of the derived gradients between different works. Blue dots represent $\tau_{\rm HCO^+} < 3$, and black dots represent $\tau_{\rm HCO^+} > 3$. The two lower limits are shifted to $R_{\rm GC} = 0.5$ kpc for better visualization. The red curve denotes the maximum likelihood fit of the blue points only; orange curves denote the 3$\sigma$ deviation from the MCMC sampling.}  \label{fig:fig2}
\end{figure}

Figure \ref{fig:fig2} shows the measured HCO$^+$/H$^{13}$CO$^+$ ratios as a function of $R_{\rm GC}$. The weighted-mean $^{12}$C/$^{13}$C ratios increase from 42$\pm$1 toward the GC to 66$\pm$5 in the solar neighborhood. To avoid an underestimation of the HCO$^+$ optical depth when $\tau$ is large, we set stricter constraints ($\tau < 3$) when fitting a gradient. We used the Markov chain Monte Carlo (MCMC) method within the $emcee$ code \citep{Foreman-Mackey2013} to sample the free parameters and the posterior probability distribution. The maximum likelihood fit of the data is shown with a red solid curve in Fig.~\ref{fig:fig2}, which gives a H$^{12}$CO$^+$/H$^{13}$CO$^+$ gradient as a function of $R_{\rm GC}$:
\begin{equation}
    \rm \frac{H^{12}CO^+}{H^{13}CO^+} = (6.4\pm1.9) \frac{{\it R}_{GC}}{kpc} + (25.9\pm10.5).
\end{equation}

\section{Analysis and discussion}\label{sec:discussion}

\subsection{The $^{12}$C/$^{13}$C ratios in the solar neighborhood}\label{sec:ratio nearby}

While the $^{12}$C/$^{13}$C ratios in the nearby molecular clouds have been derived extensively with various molecules, the most reliable tracer thought to represent the true element $^{12}$C/$^{13}$C ratios is CH$^+$ \citep{Ritchey2011}. The measured $^{12}$C/$^{13}$ ratios from our HCO$^+$ samples range from 62.0$\pm$4.4 to 139.4$\pm$38.0 within 500 pc of the solar circle ($7.6 \leq R_{\rm GC} \leq 8.6$ kpc), with a weighted mean value of 66$\pm$5. This value is in good agreement with the mean value obtained through optical $^{12}$CH$^+$/$^{13}$CH$^+$ measurements toward diffuse sight lines \citep[74.4$\pm$7.6,][]{Ritchey2011} and the $-0.9$ km\,s$^{-1}$ component of 3C~111 (Appendix \ref{sec:b}).

\subsection{The $^{12}$C/$^{13}$C ratios in the Galactic center}\label{sec:ratio in gc}

The measurements of the $^{12}$C/$^{13}$C ratio within $\sim$2 kpc of the GC are rare. This is mainly due to the high gas column density (e.g., $N_{\rm H_2} \gtrsim 10^{24}$ cm$^{-2}$), which makes the molecular transitions optically thick, even for the rare isotopologs (e.g., C$^{18}$O). \citet{Halfen2017} measured the isotopic ratios using complex organic molecules (COMs) toward Sgr B2(N) and found an average value of 24$\pm$7, which is similar to the value obtained from dense gas tracers \citep[e.g., C$^{34}$S,][]{Humire2020}. The measurements from absorption lines of CH toward Sgr B2(M) and H$_2$CO toward Sgr B2 result in $^{12}$C/$^{13}$C ratios of 15.8$\pm$2.4 \citep{Jacob2020} and 11.48$\pm$0.03 \citep{Yan2019}, respectively. 

Previous absorption observations toward the Galactic bulge have revealed a significantly high $^{12}$C/$^{13}$C ratio \citep[$\sim$60,][]{Liszt2018}. However, since these observations integrated the full velocity range, which has HCO$^+$ absorptions (the H$^{13}$CO$^+$ absorptions are much narrower), the absorption of HCO$^+$ could be contaminated with foreground diffuse gas, leading to a high value. Since the HCO$^+$ absorptions toward the dense regions (e.g., Sgr B2(N)) are completely saturated, the measured values at $R_{\rm GC} < 2$ kpc in our work are in fact the diffuse molecular components \citep[a few hundred cm$^{-3}$,][]{Gerin2017,Liszt2018}. 

The two measurements at $R_{\rm GC} < 2$ kpc from our observations have $^{12}$C/$^{13}$C ratios of 42.2$\pm$1.7 and 37.5$\pm$6.5, values which are consistent with absorption measurements of c-C$_3$H$_2$ \citep[$\sim$40 at $R_{\rm GC} < 1$ kpc,][]{Corby2018} but still larger than the other tracers in high-density ($n_{\rm H_2} > 10^5$ cm$^{-3}$) regions. Furthermore, the two lower limits of $^{12}$C/$^{13}$C at $-$181 and $-$162 km\,s$^{-1}$ of J1720-3552 are $>19$ and $>31$. These results are consistent with previous observations of CH$^+$ toward Sgr A$^*$ and Sgr B2(N) \citep[e.g., $^{12}$CH/$^{13}$CH $>$30 at $-168 \sim -150$ km\,s$^{-1}$ of Sgr A$^*$,][]{Godard2012}, which is again distinct from the previously reported low values toward Sgr B2 but supports our results. Thus, either opacity effects or fractionation may have an impact on the measured isotopic ratios in the high-density tracers (e.g., CN, and COMs), while our measurements at low density are less influenced.

\subsection{The $^{12}$C/$^{13}$C gradient and comparison with GCE models}\label{sec:gradients from different tracers}

Different tracers have been used to calculate the Galactic $^{12}$C/$^{13}$C gradient. This gradient is mostly constrained by results from $\rm ^{12}C^{18}O/^{13}C^{18}O$ \citep[e.g.,][]{Langer1990,Langer1993,Wouterloot1996,Giannetti2014}; $\rm ^{12}CN/^{13}CN$ \citep[e.g.,][]{Savage2002,Milam2005,Sun2024}; H$_2^{12}$CO/H$_2^{13}$CO \citep{Henkel1982,Yan2019}; C$^{34}$S \citep{Yan2023}; and a combined analysis from  various tracers \citep[e.g., COMs, CH;][]{Halfen2017,Jacob2020}. The gradients from different methodologies are shown with different curves in Fig.~\ref{fig:fig2}.\footnote{For different works using the same tracer, we adopted the latest gradient.} 

By comparing the gradients using different tracers, we found that the measurements in the low-density molecular clouds (HCO$^+$, $n_{\rm H_2} \sim$ a few 10$^2$ cm$^{-3}$) are higher than those from higher density environments ($n_{\rm H_2}$ $\geq$ a few 10$^4$ cm$^{-3}$), such as C$^{18}$O, H$_2$CO, CS, and CN \citep{Giannetti2014,Yan2019,Yan2023,Sun2024}. The systematically increasing trend of the $^{12}$C/$^{13}$C gradient from high-density to low-density gas may again suggest that fractionation plays a crucial role in the measured $^{12}$C/$^{13}$C ratios in molecules. Predictions from the chemical models by \citet{Colzi2020} are in agreement with this observational trend, finding that $^{12}$C/$^{13}$C ratios of HCN, HNC, and HCO$^+$ tend to increase from the higher (10$^6$ cm$^{-3}$) to the lower (10$^3$ cm$^{-3}$) densities.

\begin{figure}
\centering
\includegraphics[width=0.95\linewidth]{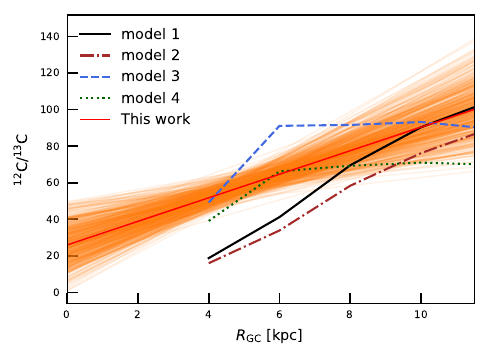}
\caption{Comparison between observations and GCE models, in which the models are taken from Table 2 in \citet{Colzi2022} with the same numbers.  \label{fig:fig3}}
\end{figure}

The comparison between the derived $^{12}$C/$^{13}$C gradient and the GCE models presented in \citet{Colzi2022} is shown in Fig.~\ref{fig:fig3}, in which the models consider different mass ranges for white dwarf progenitors and average ejected masses of $^{13}$C and $^{15}$N per nova outburst \citep{Romano2019,Romano2021}. Our result is reasonably consistent with the GCE models with a mass range of white dwarf progenitors 1--8 M$_\odot$ and higher average ejected masses of $^{13}$C (models 3 and 4). 
Given the limited range of $R_{\rm GC}$ in the current samples, future observations at $R_{\rm GC} > 10$ kpc would be exceptionally useful to constrain GCE models.

However, one should always keep in mind that the formula of the isotopic gradient strongly depends on the accuracy of the $R_{\rm GC}$, which is mostly based on the kinematic distance estimation. For instance, the estimated $R_{\rm GC}$ before and after the model by \citet{Reid2019} could differ by several kpc \citep{Sun2024}. The large scatter of $^{12}$C/$^{13}$C at $R_{\rm GC}$ between 4 kpc to 8 kpc in both our work and the literature may suggest that our understanding of the Milky Way rotational model or inhomogeneous mixing of the elements may also contribute to the scatter.

\subsection{Chemical effects on the measured $^{12}$C/$^{13}$C ratios}\label{sec:chemical effects}

The isotopic exchange reactions could be one of the main processes that lead to different isotopic ratios in HCO$^+$.
In this section, we consider the isotopic exchange reaction that produces H$^{13}$CO$^+$ \citep{Langer1978,Roueff2015,Colzi2020}\footnote{The forward reaction only has high efficiency ($\geq$ 50\%) in cold environments (e.g., $T_{\rm k} < 30$ K).}:
\begin{align}
\rm ^{13}C^+ + CO \rightleftharpoons &\ \rm ^{13}CO + C \tag{R1},\label{r:R1}\\
\rm ^{13}CO + HCO^+ \rightleftharpoons &\ \rm H^{13}CO^+ + CO \tag{R2}.
\label{r:R2}
\end{align}
We considered the main formation pathways of H$^{13}$CO$^+$ in low-to-intermediate density gas \citep{vanDishoeck1988,Luo2023b}:
\begin{align}
\rm ^{13}CO^+ + H_2 \rightarrow &\ \rm H^{13}CO^+ + H \tag{R3},\label{r:R3}\\
\rm ^{13}C^+ + H_2O \rightarrow &\ \rm H^{13}CO^+ + H \tag{R4},\label{r:R4}\\
\rm ^{13}CH + O \rightarrow &\ \rm H^{13}CO^+ + e \tag{R5}.
\label{r:R5}
\end{align}
The destruction of HCO$^+$ (and H$^{13}$CO$^+$) is dominated by electrons:
\begin{equation}
    \rm HCO^+ + e^- \rightarrow CO + H \tag{R6}.
    \label{r:R6}
\end{equation} 
Thus, the isotopic exchange reaction would influence the HCO$^+$/H$^{13}$CO$^+$ ratio only if the right side reaction \ref{r:R2} is comparable to reactions \ref{r:R3} $\sim$ \ref{r:R5}. The reaction rate of the above reaction is $2.6\times10^{-10}\times(T/300)^{-0.4}$ cm$^{3}$\,s$^{-1}$ \citep{Roueff2015}. The typical abundance of CO$^+$ is $10^{-10} \sim 10^{-9}$\citep{Stauber2009,Trevino-Morales2016}, the abundance of HCO$^+$ is a few $10^{-9}$\citep{Lucas1996,Gerin2019,Luo2020}, and the abundance of CH is $3\times10^{-8}$ \citep{Liszt2002,Sheffer2008,Tang2021,Luo2023a}. Even if we only consider reaction \ref{r:R3} (the reaction rate k = $7.5\times10^{-10}$ cm$^{3}$\,s$^{-1}$; \citet{McElroy2013}) and scale the above abundance with $^{12}$C/$^{13}$C = 70, the formation of H$^{13}$CO$^+$ through reaction \ref{r:R3} would overwhelm reaction \ref{r:R2} by over four orders of magnitude at the typical gas temperature of $\sim$50 K \citep{Snow2006}. Therefore, fractionation would have little impact on the measured HCO$^+$/H$^{13}$CO$^+$ ratios in low-density gas. This is also consistent with various chemical modelings of carbon fractionation, in which HCO$^+$/H$^{13}$CO$^+$ can overall best represent the original $^{12}$C/$^{13}$C element ratio in low-to-intermediate density gas \citep{Szucs2014,Roueff2015,Colzi2020,Sipila2023}. 

In dense regions, CO becomes the precursor of HCO$^+$ \citep{Dalgarno2006,Indriolo2012,Bisbas2015}:
\begin{equation}
    \rm CO + H_3^+ \rightarrow HCO^+ + H_2 \tag{R7}.
    \label{r:R7}
\end{equation} 
Since reaction \ref{r:R6} is still the dominant formation channel of CO even in high-density clouds \citep{Luo2023a}, the fractionation of CO in the dense cloud could greatly impact the isotopic ratio in HCO$^+$ only if reaction \ref{r:R7} is comparable to reaction \ref{r:R6}. However, the gas density that will significantly alter the $^{12}$C/$^{13}$C ratio measured in HCO$^+$ would be above 10$^4$ cm$^{-3}$ \citep{Sipila2023}.
The highest column density of HCO$^+$ in Table \ref{table:tab1} is $\sim 3.4\times10^{13}$ cm$^{-2}$, corresponding to an H$_2$ column density of $10^{22}$ cm$^{-2}$, assuming a constant HCO$^+$ abundance \citep[$3\times10^{-9}$,][]{Lucas1996,Liszt2016}\footnote{However, since the integrated velocity range is over 20 km/,s$^{-1}$ and contains a few velocity components, the actual column density of each component should be a few fractions of the total.} Therefore, most components are not expected to trace such a high-density regime and will be less influenced by the fractionation effect. Nevertheless, both detailed chemical modeling (Colzi et al. in prep) and the constraints of gas volume density and temperature through multiple lines in the future are necessary to determine where and how fractionation impacts the isotopic ratios.

\section{Conclusion}\label{sec:conclusion}

In this work, we have performed high sensitivity absorption line observations toward strong continuum sources (including quasars and UCH\,{\sc ii} regions). We derived $^{12}$C/$^{13}$C ratios from HCO$^+$ absorption and $R_{\rm GC}$ according to the latest parallax-based distance calculation. Our main conclusions are as follows:
\begin{enumerate}
\item The derived $^{12}$C/$^{13}$C gradient from HCO$^+$ absorption measurements is (6.4$\pm$1.9)$R_{\rm GC}$/kpc+(25.9$\pm$10.5), which is reasonably consistent with current GCE models.

\item The derived weighted mean $^{12}$C/$^{13}$C ratio in the solar neighborhood is 66$\pm$5, which is consistent with those measured from CH$^+$ (74.4$\pm$7.6).

\item Our measurements toward the GC are two to three times higher than those measured with dense gas tracers toward Sgr B2 (11$\sim$24). Nevertheless, our results are supported by the CH$^+$ observations toward Sgr A$^*$ and Sgr B2(N) ($>$30) as well as absorption measurements of c-C$_3$H$_2$ ($\sim$40). 

\item The discrepancy between our method and those from dense gas tracers suggests that opacity effects and fractionation may have a larger impact on the dense gas tracers in high-density regions.

\end{enumerate}

We highlight the use of absorption lines to measure the isotopic ratios with interferometry observations, which are less affected by optical depth, beam dilution, and chemical fractionation. Future large samples toward the GC and the anti-center directions may provide more constraints on the Galactic $^{12}$C/$^{13}$C gradient as well as the GCE models.


\begin{acknowledgements}
We are grateful to the anonymous referee for the thoughtful comments and suggestions that greatly improved the clarity of our work, especially the suggestion of including the non-circular motion to estimate the distance near the GC. We thank Zhiyu Zhang, J$\rm {\acute{e}}$r$\rm {\hat{o}}$me Pety, and Michel Gu$\rm {\acute{e}}$lin for their useful comments, Donatella Romano for providing the GCE models, and the staffs at IRAM for carrying out the NOEMA observations and reducing the data. L. C acknowledges support from the grant No. PID2022-136814NB-I00 by the Spanish Ministry of Science, Innovation and Universities/State Agency of Research MICIU/AEI/10.13039/501100011033 and by ERDF, UE. Tie Liu acknowledges the supports by the National Key R\&D Program of China (No. 2022YFA1603100), National Natural Science Foundation of China (NSFC) through grants No.12073061 and No.12122307, and the Tianchi Talent Program of Xinjiang Uygur Autonomous Region. D. L. is a New Cornerstone investigator. N.-Y. Tang is sponsored by the University Annual Scientific Research Plan of Anhui Province (No. 2023AH030052, No. 2022AH010013), the China Manned Space Program through its Space Application System,  Zhejiang Lab Open Research Project (No. K2022PE0AB01). 
This paper makes use of the following ALMA data: ADS/JAO.ALMA\#2022.1.01438.S and ADS/JAO.ALMA\#2019.1.00685.S. ALMA is a partnership of ESO (representing its member states), NSF (USA) and NINS (Japan), together with NRC (Canada), MOST and ASIAA (Taiwan), and KASI (Republic of Korea), in cooperation with the Republic of Chile. The Joint ALMA Observatory is operated by ESO, AUI/NRAO and NAOJ. This work is based on observations carried out under project number W20BB with the IRAM NOEMA Interferometer. IRAM is supported by INSU/CNRS (France), MPG (Germany) and IGN (Spain).
\end{acknowledgements}

%
%

\bibliographystyle{aa}
\bibliography{example} 



\begin{appendix} 

\section{Absorption spectra toward all sources}\label{sec:a}

\begin{figure*}[!hbt]
\centering
\includegraphics[width=0.85\linewidth]{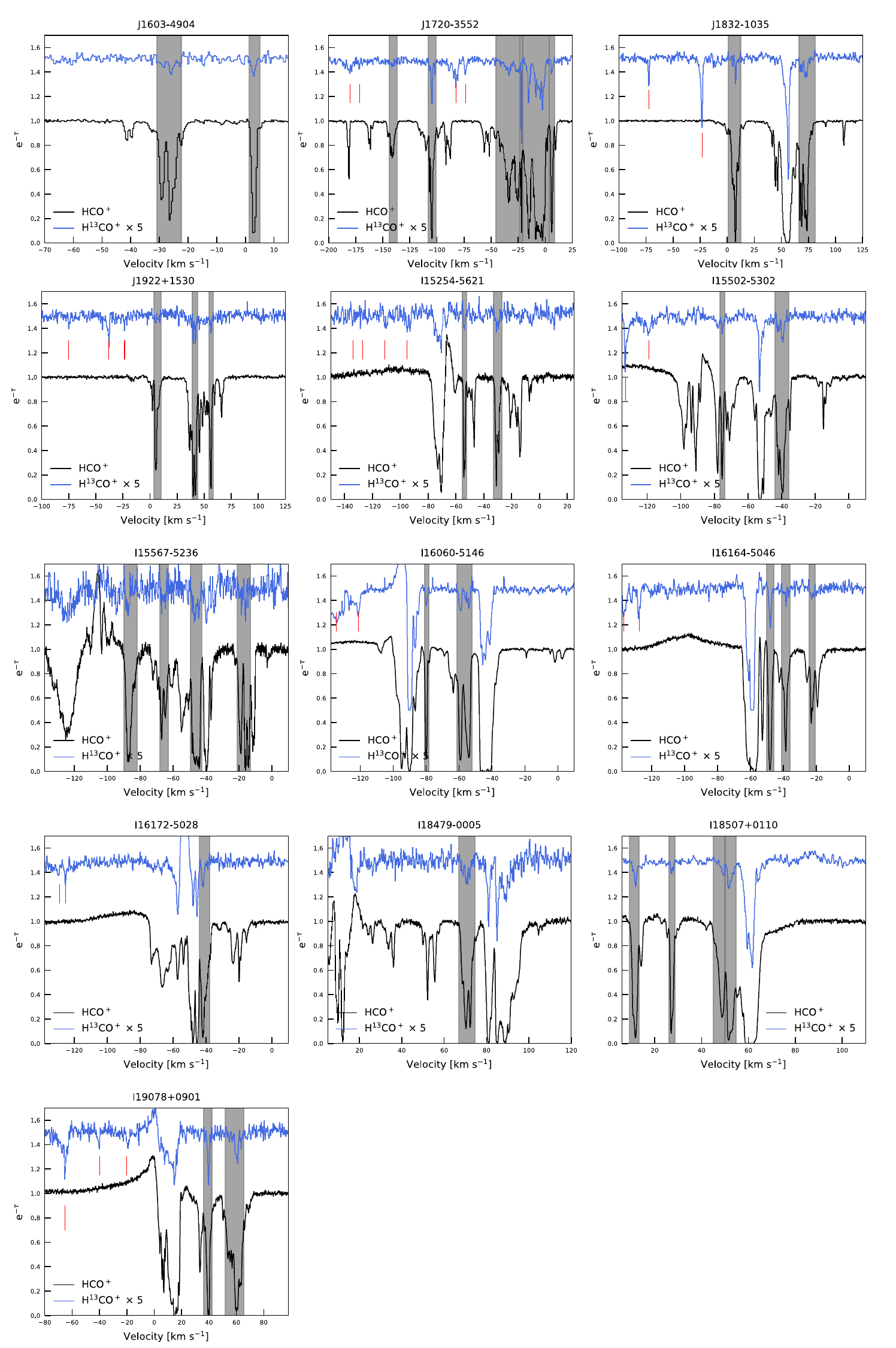}
\caption{Same as Fig.~\ref{fig:fig1} but for other sources. Red vertical lines denote HCO absorption. \label{fig:allspec}}
\end{figure*}

The normalized absorption profiles of 14 sources are shown in Fig.~\ref{fig:allspec}. Note that for the UCH\,{\sc ii} sight lines, the H40$\alpha$ emissions are usually associated with UCH\,{\sc ii} regions \citep{Liu2020} (e.g., $-150 \sim -80$ km\,s$^{-1}$ toward I15254+5621), this is not the bad baseline. Furthermore, since the emission from the compact structures around UCH\,{\sc ii} regions could have a high excitation temperature, the spectra may still show emission in conjunction with absorption. We avoid these velocity ranges in our calculation. Future high angular resolution and high sensitivity observations may help reconstruct both the emission and absorption profiles around UCH\,{\sc ii} regions.

\section{Gaussian decomposition of absorption profiles toward 3C~111}\label{sec:b}

\begin{figure*}
\centering
\includegraphics[width=0.98\linewidth]{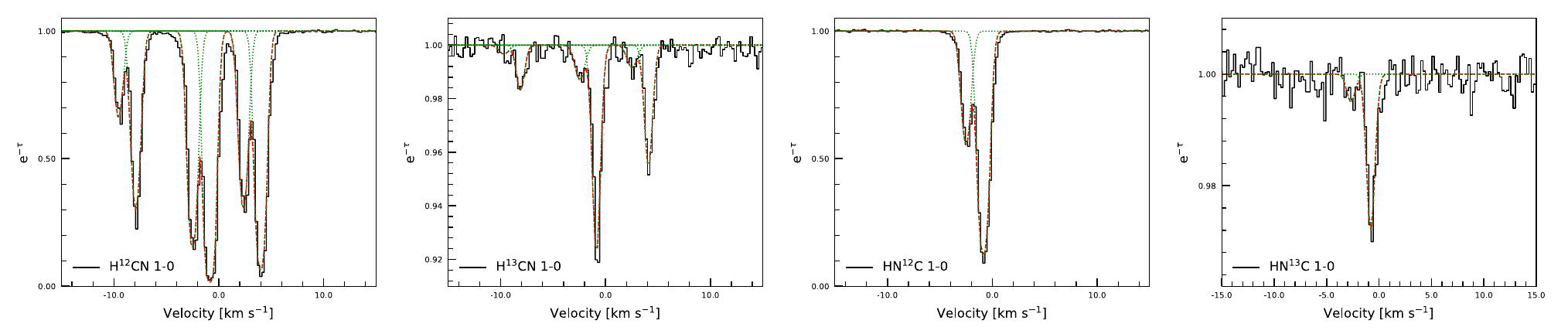}
\caption{Gaussian decomposition of normalized absorption profiles of HCN, H$^{13}$CN, HNC, and H$^{13}$NC toward 3C~111. Red dashed curves denote the fitting results, and green dotted lines denote each Gaussian component. \label{fig:3c111}}
\end{figure*}

Absorption profiles and Gaussian decomposition of H$^{12}$CN, H$^{13}$CN, HN$^{12}$C, and HN$^{13}$C {\it J}=1-0 transitions toward 3C~111 are shown in Fig.~\ref{fig:3c111}, in which the spectra of H$^{12}$CN and H$^{13}$CN are hyperfine transitions. The Gaussian decomposition is performed with $curve\_fit$ package in $scipy$ using the Levenberg-Marquardt algorithm, and the resultant optical depths and linewidth ($\Delta V$) are shown in Table \ref{table:tab1}. We set constraints when fitting hyperfine transitions that all hyperfine lines have 1) the same linewidth ($\Delta V$) and 2) the same velocity offset with respect to $V_{\rm lsr}$.

There are two velocity components ($-0.9$ km\,s$^{-1}$ and $-2.5$ km\,s$^{-1}$) in front of 3C~111, the 3D extinction indicates that the cloud is within 300 pc from the Sun \citep{Lucas1998}. The gas volume density at $-0.9$ km\,s$^{-1}$ is $n_{\rm H_2} = 398\pm22$ cm$^{-3}$ (Luo et al. in prep), and $n_{\rm H_2}$ is supposed to be much lower at $-2.5$ km\,s$^{-1}$ than at $-0.9$ km\,s$^{-1}$ \citep{Lucas1998}. However, the HCO$^+$ absorption profile is almost saturated toward 3C~111 and we cannot get a good Gaussian decomposition. Thus, we use HCN and HNC to derive the $^{12}$C/$^{13}$C ratio. 
The optical depth ratios of the H$^{12}$CN hyperfine transitions are 1:2.3:3.5 and 1:2.9:4.6 at $-$0.9 km\,s$^{-1}$ and $-$2.5 km\,s$^{-1}$, respectively. While the latter represents the intrinsic line ratio (1:3:5), the former deviates from it by 30\%, indicating that the optical depths of the H$^{12}$CN (F=1--1 and 2--1) at $-$0.9 km\,s$^{-1}$ are underestimated. As an alternative, we can fix the optical depth ratios at 1:3:5 instead of treating them as free parameters to obtain a more accurate value of the column density.

The $^{12}$C/$^{13}$C ratio at $-0.9$ km\,s$^{-1}$ derived from H$^{12}$CN/H$^{13}$CN is 74$\pm$3, which is 30\% higher than that previously obtained by \citet{Lucas1998}. Measurement from HN$^{12}$C/HN$^{13}$C (62$\pm$6) also show similar result. 
The $^{12}$C/$^{13}$C ratio at $-2.5$ km\,s$^{-1}$ component derived from H$^{12}$CN/H$^{13}$CN is 138$\pm$28, which is $\sim$2 times higher than that at $-0.9$ km\,s$^{-1}$. As suggested by \citet{Lucas1998}, this could be a particular case, where the $^{13}$CO has been enriched and fractionation leads to a deficit of $^{13}$C in less abundant carbon carriers. Chemical models that consider the isotopic fractionation also predict a higher isotopic ratio of H$^{12}$CN/H$^{13}$CN than that from CO or HCO$^+$ \citep{Roueff2015,Colzi2020}. Our detection of HN$^{13}$C at $-2.5$ kms\,s$^{-1}$ (S/N of $\int \tau dv$ $\sim$ 4) is consistent with the hypothesis (HN$^{12}$C/HN$^{13}$C = $95\pm39$). 

\begin{table*}
\caption{Optical depths and line widths toward 3C~111.}
\label{table:tab1}   
\centering          
\begin{tabular}{c c c c c c}     
\hline\hline  
{Molecules}& \multicolumn{2}{c}{$-0.9$ km\,s$^{-1}$} & \multicolumn{2}{c}{$-2.5$ km\,s$^{-1}$} \\
\cline{2-3} \cline{4-5}
 & $\tau$ & {$\Delta V$ (km\,s$^{-1}$)}  & {$\tau$} & {$\Delta V$ (km\,s$^{-1}$)}  \\
\hline
H$^{12}$CN & 1.21$\pm$0.02\tablefootmark{a} & 0.94$\pm$0.01 &  0.42$\pm$0.02\tablefootmark{a} & 0.93$\pm$0.01  \\
H$^{13}$CN & 0.082$\pm$0.002 & 0.90$\pm$0.03  & 0.013$\pm$0.002 & 1.1$\pm$0.2  \\
HN$^{12}$C & 2.14$\pm$0.03 & 1.03$\pm$0.01  & 0.595$\pm$0.008 & 1.0$\pm$0.2  \\
HN$^{13}$C & 0.028$\pm$0.002 & 0.94$\pm$0.07 & 0.005$\pm$0.002 & 0.95$\pm$0.4\\
\hline                  
\end{tabular}
\tablefoot{\tablefoottext{a}{Optical depth of {\it J}=1-0, {\it F}=0-1 transition.}}
\end{table*}

\section{The source properties}\label{sec:c}

The derived source properties from Sect.~\ref{sec:results} are shown in Table \ref{table:tab2}.

\begin{table*}
\caption{Source name, velocity, galactocentric distance, integrated optical depths, isotopic ratios, and RMS of the absorption spectra.}             
\label{table:tab2}      
\centering          
\begin{tabular}{c c c c c c c}     
\hline\hline       
Source & $V_{\rm lsr}$ & $R_{\rm GC}$ & $\int \tau_{\rm H^{12}CO^+} d\upsilon$ & $\int \tau_{\rm H^{13}CO^+} d\upsilon$ & $N_{\rm H^{12}CO^+}$/$N_{\rm H^{13}CO^+}$ & RMS \\
\cline{2-7}
 & km\,s$^{-1}$ & kpc & km\,s$^{-1}$ & km\,s$^{-1}$ & & $10^{-3}$\\
\hline                    
I15254-5621 & -54.0$\pm$0.7 & 6.0$^{+0.2}_{-0.2}$ & 1.944$\pm$0.017 & 0.019$\pm$0.007 & 101.8$\pm$37.4 & 11.7 \\ 
I15254-5621 & -30.4$\pm$0.3 & 6.8$^{+0.2}_{-0.2}$ & 3.166$\pm$0.023 & 0.038$\pm$0.010 & 82.1$\pm$21.5 & 11.7 \\ 
I15502-5302 & -75.3$\pm$0.8 & 5.1$^{+0.1}_{-0.2}$ & 1.787$\pm$0.013 & 0.038$\pm$0.003 & 46.0$\pm$4.0 & 9.0 \\ 
I15502-5302 & -40.0$\pm$0.3 & 6.3$^{+0.2}_{-0.2}$ & 8.526$\pm$0.048 & 0.127$\pm$0.006 & 65.1$\pm$3.0 & 9.0 \\ 
I15567-5236 & -86.6$\pm$2.3 & 4.7$^{+0.1}_{-0.2}$ & 7.101$\pm$0.130 & 0.072$\pm$0.026 & 96.6$\pm$35.0 & 30.9 \\ 
I15567-5236 & -65.9$\pm$1.4 & 5.3$^{+0.2}_{-0.1}$ & 3.254$\pm$0.047 & 0.060$\pm$0.021 & 52.9$\pm$18.2 & 30.9 \\ 
I15567-5236 & -46.0$\pm$1.6 & 6.0$^{+0.2}_{-0.2}$ & 12.476$\pm$0.307 & 0.197$\pm$0.024 & 61.6$\pm$7.7 & 30.9 \\ 
I15567-5236 & -16.4$\pm$1.0 & 7.3$^{+0.2}_{-0.2}$ & 11.112$\pm$0.489 & 0.091$\pm$0.026 & 118.4$\pm$34.0 & 30.9 \\ 
I16060-5146 & -80.0$\pm$0.5 & 4.7$^{+0.2}_{-0.2}$ & 2.381$\pm$0.011 & 0.037$\pm$0.003 & 62.6$\pm$4.7 & 4.5 \\ 
I16060-5146 & -56.7$\pm$0.2 & 5.5$^{+0.2}_{-0.2}$ & 10.094$\pm$0.022 & 0.159$\pm$0.005 & 62.0$\pm$2.0 & 4.5 \\ 
I16164-5046 & -48.0$\pm$1.5 & 5.7$^{+0.2}_{-0.2}$ & 5.727$\pm$0.126 & 0.106$\pm$0.005 & 52.5$\pm$2.7 & 8.9 \\ 
I16164-5046 & -38.7$\pm$0.3 & 6.1$^{+0.2}_{-0.2}$ & 2.976$\pm$0.016 & 0.043$\pm$0.005 & 67.5$\pm$8.3 & 8.9 \\ 
I16164-5046 & -22.5$\pm$0.2 & 6.8$^{+0.2}_{-0.2}$ & 1.676$\pm$0.009 & 0.032$\pm$0.004 & 50.9$\pm$7.1 & 8.9 \\ 
I16172-5028 & -41.3$\pm$0.4 & 5.9$^{+0.2}_{-0.2}$ & 6.853$\pm$0.043 & 0.096$\pm$0.006 & 69.5$\pm$4.4 & 8.0 \\ 
I18479-0005 & 70.8$\pm$0.6 & 5.2$^{+0.2}_{-0.1}$ & 6.478$\pm$0.036 & 0.113$\pm$0.012 & 55.7$\pm$6.1 & 11.7 \\ 
I18507+0110 & 11.6$\pm$0.2 & 7.6$^{+0.2}_{-0.2}$ & 4.714$\pm$0.056 & 0.074$\pm$0.006 & 62.0$\pm$4.8 & 11.4 \\ 
I18507+0110 & 27.3$\pm$0.4 & 6.9$^{+0.2}_{-0.2}$ & 2.579$\pm$0.027 & 0.029$\pm$0.005 & 87.7$\pm$14.2 & 11.4 \\ 
I18507+0110 & 48.3$\pm$0.4 & 6.0$^{+0.2}_{-0.2}$ & 3.619$\pm$0.021 & 0.043$\pm$0.006 & 82.5$\pm$12.1 & 11.4 \\ 
I18507+0110 & 52.2$\pm$0.9 & 5.9$^{+0.1}_{-0.2}$ & 8.472$\pm$0.109 & 0.123$\pm$0.006 & 67.3$\pm$3.5 & 11.4 \\ 
I19078+0901 & 39.4$\pm$2.8 & 6.7$^{+0.2}_{-0.2}$ & 8.199$\pm$0.404 & 0.154$\pm$0.008 & 52.0$\pm$3.8 & 11.4 \\ 
I19078+0901 & 59.6$\pm$0.6 & 6.0$^{+0.2}_{-0.1}$ & 15.748$\pm$0.106 & 0.188$\pm$0.012 & 81.6$\pm$5.3 & 11.4 \\ 
J1603-4904 & -27.0$\pm$0.2 & 6.7$^{+0.2}_{-0.2}$ & 5.195$\pm$0.026 & 0.072$\pm$0.010 & 70.4$\pm$9.6 & 6.0 \\ 
J1603-4904 & 2.8$\pm$0.1 & 8.5$^{+0.2}_{-0.2}$ & 3.352$\pm$0.043 & 0.036$\pm$0.007 & 91.6$\pm$16.9 & 6.0 \\ 
J1720-3552 & -181.4$\pm$1.6 & 1.37$^{+0.04}_{-0.04}$\tablefootmark{a} & 0.927$\pm$0.005 & $<$0.047\tablefootmark{b} & $>19$ & 3.4 \\
J1720-3552 & -161.9$\pm$2.4 & 1.37$^{+0.04}_{-0.04}$\tablefootmark{a} & 0.679$\pm$0.007 & $<$0.021 & $>31$ & 3.4 \\
J1720-3552 & -140.9$\pm$1.0 & 1.37$^{+0.04}_{-0.04}$\tablefootmark{a} & 1.426$\pm$0.007 & 0.037$\pm$0.006 & 37.5$\pm$6.5.5 & 3.4 \\ 
J1720-3552 & -104.6$\pm$1.3 & 1.3$^{+0.0}_{-0.0}$ & 6.219$\pm$0.056 & 0.144$\pm$0.005 & 42.2$\pm$1.7 & 3.4 \\ 
J1720-3552 & -31.7$\pm$0.1 & 4.2$^{+0.2}_{-0.2}$ & 10.925$\pm$0.019 & 0.227$\pm$0.011 & 46.8$\pm$2.3 & 3.4 \\ 
J1720-3552 & -22.0$\pm$0.2 & 5.0$^{+0.2}_{-0.2}$ & 2.321$\pm$0.018 & 0.178$\pm$0.004 & 12.7$\pm$0.3 & 3.4 \\ 
J1720-3552 & -7.9$\pm$0.0 & 6.8$^{+0.3}_{-0.3}$ & 34.042$\pm$0.124 & 0.784$\pm$0.012 & 42.3$\pm$0.7 & 3.4 \\ 
J1720-3552 & 6.3$\pm$0.1 & 6.9$^{+0.1}_{-0.1}$ & 3.447$\pm$0.021 & 0.043$\pm$0.005 & 78.6$\pm$9.5 & 3.4 \\ 
J1832-1035 & 7.1$\pm$0.2 & 7.6$^{+0.2}_{-0.2}$ & 6.393$\pm$0.136 & 0.045$\pm$0.013 & 139.4$\pm$41.9 & 3.7 \\ 
J1832-1035 & 70.9$\pm$0.3 & 4.3$^{+0.1}_{-0.1}$ & 10.348$\pm$0.032 & 0.140$\pm$0.018 & 72.1$\pm$9.2 & 3.7 \\ 
J1851+0035 & 14.5$\pm$0.2 & 7.5$^{+0.2}_{-0.2}$ & 9.555$\pm$0.077 & 0.093$\pm$0.021 & 99.6$\pm$22.7 & 5.5 \\ 
J1851+0035 & 32.5$\pm$0.4 & 6.6$^{+0.2}_{-0.2}$ & 1.971$\pm$0.017 & 0.023$\pm$0.011 & 82.6$\pm$40.8 & 5.5 \\ 
J1851+0035 & 55.6$\pm$1.1 & 5.8$^{+0.2}_{-0.2}$ & 11.614$\pm$0.162 & 0.136$\pm$0.019 & 83.4$\pm$11.8 & 5.5 \\ 
J1922+1530 & 5.8$\pm$0.1 & 8.0$^{+0.2}_{-0.2}$ & 3.135$\pm$0.024 & 0.025$\pm$0.010 & 123.7$\pm$47.7 & 6.9 \\ 
J1922+1530 & 40.9$\pm$1.1 & 6.8$^{+0.2}_{-0.2}$ & 8.187$\pm$0.156 & 0.116$\pm$0.009 & 68.8$\pm$5.3 & 6.9 \\ 
J1922+1530 & 55.9$\pm$1.1 & 6.3$^{+0.2}_{-0.1}$ & 3.238$\pm$0.044 & 0.042$\pm$0.007 & 75.2$\pm$13.4 & 6.9 \\   
\hline                  
\end{tabular}
\tablefoot{\tablefoottext{a}{Estimation through tilted-disk model near the GC (see Sect.~\ref{sec:12_13c_gradient}).}
\tablefoottext{b}{H$^{13}$CO$^+$ is contaminated with HCO, thus, it is an upper limit.}}
\end{table*}

\section{The Molecular transitions}\label{sec:d}

The molecular transitions mentioned in this work are taken from the CDMS database \citep{Muller2001,Muller2005}, which are listed in Table \ref{tab:transitions}.

\begin{table}
\caption{Molecular transitions mentioned in this work.}  
\label{tab:transitions}
\centering          
\begin{tabular}{c c c c}     
\hline\hline  
{Transition}& {$\nu$} & {$\left | \mu_{\rm lu} \right |^2$} & $E_{\rm u}$\\
\cline{2-4} 
 & {MHz} & Debye$^2$  & K   \\
 \hline
H$^{12}$CO$^+$ (1-0) & 89188.5247 & 15.21 & 4.28\\
H$^{13}$CO$^+$ (1-0) & 86754.2884 & 15.21 & 4.16\\
H$^{12}$CO (1/2-1/2, F=1-1) & 86777.46 & 1.86 & 4.18\\
H$^{12}$CN (1-0, F=1-1) & 88630.4156 & 8.9 & 4.25\\
H$^{12}$CN (1-0, F=2-1) & 88631.8475 & 8.9 & 4.25\\
H$^{12}$CN (1-0, F=0-1) & 88633.9357 & 8.9 & 4.25\\
H$^{13}$CN (1-0, F=1-1) & 86338.7352 & 8.9 & 4.14\\
H$^{13}$CN (1-0, F=2-1) & 86340.1666 & 8.9 & 4.14\\
H$^{13}$CN (1-0, F=0-1) & 86342.2543 & 8.9 & 4.14\\
HN$^{12}$C (1-0) & 90663.568 & 9.3 & 4.35\\
HN$^{13}$C (1-0) & 87090.85 & 7.28 & 4.18\\
\hline                  
\end{tabular}
\end{table}

\section{The unusual low H$^{12}$CO$^+$/H$^{13}$CO$^+$ toward J1720-3552}\label{sec:e}

The $-22$ km\,s$^{-1}$ velocity component was recognized as an independent component in the H$^{13}$CO$^+$ line profiles when we calculated the H$^{12}$CO$^+$/H$^{13}$CO$^+$ ratios. The derived isotopic ratio is more than three times lower than the components toward the GC; however, the reason is still unclear. If this low value is caused by chemical fractionation, the gas component at $-22$ km\,s$^{-1}$ must have a high density and compact size ($\leq$0.01 pc at a distance of $\sim$ 3 kpc; otherwise, we would see emission lines around the continuum source). It should be surrounded by low-density envelopes since the H$^{12}$CO$^+$/H$^{13}$CO$^+$ ratios toward the neighboring velocity components are much higher. Future high-$J$ molecular line observations may help reveal the nature of this low value.

\end{appendix}
%
%
\end{document}